\documentclass[12pt,a4paper]{article}

\usepackage{amsfonts}
\usepackage{amsmath}
\usepackage{amssymb}
\usepackage{latexsym}

\usepackage[dvips]{graphicx}

\numberwithin{equation}{section}

\newcommand{\lanln}[1]{$\langle$\texttt{arXiv:#1}$\rangle$}

\newcommand{\BbbR}{\mathbb{R}}

\title{Then again, how often does the Unruh-DeWitt detector click if we switch it carefully?}
\author{Alejandro Satz\thanks{pmxas3@nottingham.ac.uk}
\\
\noalign{\vspace{3ex}}
\small{\it School of Mathematical Sciences,
University of Nottingham,}\\
\small{\it Nottingham NG7 2RD, UK}
\\
\\
\\
\noalign{\vspace{1ex}}
}

\date{}

\begin{document}

\maketitle

\begin{abstract}

The transition probability in first-order perturbation theory for an Unruh-DeWitt detector coupled to a massless scalar field in Minkowski space is calculated. It has been shown recently that the conventional $i\epsilon$ regularisation prescription for the correlation function leads to non-Lorentz invariant results for the transition rate, and a different regularisation, involving spatial smearing of the field, has been advocated to replace it. We show that the non-Lorentz invariance arises solely from the assumption of sudden switch-on and switch-off of the detector, and that when the model includes a smooth switching function the results from the conventional regularisation are both finite and Lorentz invariant. The sharp switching limit of the model is also discussed, as well as the falloff properties of the spectrum for large frequencies.

\end{abstract}

\section{Introduction}

A common way to probe the physics of quantum fields when accelerated observers or curved backgrounds are involved is via the use of particle detectors. This approach, initiated by Unruh \cite{unruh} and DeWitt \cite{deWitt}, attempts to define the particle content of the field by the transitions induced between the discrete levels of a simple quantum system (``atom'' or ``detector'') which is weakly coupled to the field. The best known application of this technique is Unruh's original discovery that the excitation spectrum is thermal for a uniformly accelerated detector in Minkowski space \cite{unruh}; similar results are found for detectors at rest in exterior Schwarzschild space \cite{hawking} and for inertial detectors in de Sitter space \cite{gibb-haw:dS}. 

In all these cases, the detector's trajectory is stationary (meaning that it is an orbit of a timelike Killing vector) and the excitation probability over the complete trajectory is formally infinite. It is the transition probability per unit time or transition rate that is well-defined and displays the thermal properties. However, for a trajectory with arbitrarily varying acceleration the transition rate has to be defined as the (proper) time-derivative of the probability of the detector having made a transition up to a certain time, and this definition involves considerable subtleties.

Schlicht \cite{schlicht} has shown that, even for the uniformly accelerated trajectory in Minkowski space, the transition rate defined as instantaneous proper time derivative of the transition probability has unphysical properties if calculated using the standard $i \epsilon$ regularisation for the massless scalar field's Wightman function. He proposed to replace the conventional regularisation by one based on a finite spatial extension for the detector, finding that this ``profile regularisation'' gives physically acceptable results \cite{schlicht,Schlicht:thesis}. P. Langlois has generalised successfully the profile regularisation setting to a variety of cases such as the massive scalar field, the massless Dirac field and some examples of curved spaces \cite{Langlois,Langlois-thesis}. A further extension of Schlicht's results is archieved in \cite{us}, where it is shown: firstly, that the transition rate defined with the $i\epsilon$ reguilarisation is non-Lorentz invariant for all non-inertial trajectories, and secondly, that its definition by Schlicht's procedure can be rewritten as a compact expression that involves no regulators or limits, and which follows as well (under certain technical assumptions) from a more general class of spatial profiles than the one considered by Schlicht.

However, the precise reason why the usual $i \epsilon$ regularisation fails was addressed at length in none of \cite{schlicht,Schlicht:thesis,Langlois,Langlois-thesis,us}. After all, it is a theorem \cite{kay-wald} that the $i \epsilon$ regulated Wightman function defines a distribution suitable for integration against functions on $M^4 \times M^4$, and this distribution can be pulled back to a well-defined distribution on the worldline of the pointlike detector \cite{Junker}. The catch pointed out in \cite{us} is that the functions integrated against have to be smooth, while the definition of the instantaneous excitation rate requires a sharp cut-off in the integral at the time of measurement. But it was not analysed how exactly could the sharpness of the cut-off affect the results to make them unphysically Lorentz-noninvariant. To provide such an analysis is the goal of the present paper.

The main result is that if the detector is turned on and off using smooth switching functions, the non-Lorentz invariant terms in the transition probability cancel out and the results are fully physical. The transition rate requires a sharp turn-off to be precisely defined, but we can show that if the switch-on and switch-off times are short compared to the total detection time, the derivative of the transition probability with respect to this total detection time is well-defined and its leading behaviour in the short switching time limit agrees with the general expression found from the spatial profile models in \cite{us}.

The paper is structured as follows: In section 2 we revisit the definition of the Unruh-DeWitt particle detector model in Minkowski space and how its transition probability and transition rate are defined. We also summarize the results of \cite{schlicht} and \cite{us} as to how the Lorentz-noninvariance appears and how it is eliminated by the spatial profile regularisation. In section 3 we present a model in which the detector is turned on and off smoothly, and we verify that the conventional regularisation for the transition probability is now fully Lorentz invariant. In section 4 we consider the limit in which the switching functions approach step functions, and we show that the results of \cite{us} can be recovered by a suitable definition of the transition rate in this limit. Section 5 discusses general properties of the response function, in particular its asymptotic behaviour for large frequencies. The results are summarised and discussed in section 6.

We use signature $(- + + +)$  for the Minkowski metric and units in which $c=\hbar=1$. Boldface letters denote spatial 
three-vectors and sans-serif letters spacetime four-vectors. The Euclidean scalar product of three-vectors $\mathbf{k}$ and $\mathbf{x}$ is denoted by $\mathbf{k}\cdot \mathbf{x}$, and the Minkowski scalar product of four-vectors $\mathsf{k}$ and $\mathsf{x}$ is denoted by $\mathsf{k}\cdot \mathsf{x}$. $O(x)$ denotes a quantity for which $O(x)/x$ is bounded as $x\to0$.

\section{Particle detectors: a brief summary}

We consider a detector consisting of an idealised atom with two energy levels, $\vert0\rangle_d$ and $\vert1\rangle_d$, with associated energy eigenvalues $0$ and $\omega$. The detector is following a trajectory $\mathsf{x}(\tau)$ in Minkowski space, parametrized by its proper time $\tau$. We introduce a coupling to a real, massless scalar field $\phi$ by modelled by the interaction Hamiltonian:

\begin{equation}\label{Hint}
 H_{\mathrm{int}}=c \chi(\tau) \mu(\tau)\phi 
\bigl(\mathsf{x}(\tau) \bigr) \,.
\end{equation}
Here $c$ is a coupling constant assumed small, $\mu(\tau)$ is the atom's monopole moment operator and $\chi(\tau)$ is a smooth switching function of compact support, positive during the interaction and vanishing before and after the interaction. If the detector is prepared in state $\vert0\rangle_d$ and the field is in a state $\vert A\rangle$ before the interaction, it may happen that after the interaction the detector is found in state $\vert1\rangle_d$. This transition can heuristically be interpreted as the absortion of a particle of energy $\omega$ from the field, if $\omega >0$, or an emission if $\omega <0$. The probability of the transition can be calculated in first-order perturbation theory and reads \cite{Junker, byd, wald-smallbook}:

\begin{equation}
\label{total-probability}
P(\omega)
=
c^2
\, 
{\bigl\vert
{}_d\langle0\vert\mu(0)\vert1\rangle_d\bigr\vert}^2
F(\omega)
\ , 
\end{equation}
where the response function 
$F(\omega)$ is given by 
\begin{equation}
\label{defresponse}
F(\omega)= \int_{-\infty}^{\infty}
\mathrm{d}\tau' \int_{-\infty}^{\infty}\mathrm{d}\tau'' 
\, 
\mathrm{e}^{-i\omega(\tau'-\tau'')}
\, 
\chi(\tau')\chi(\tau'')
\, 
\langle A\vert 
\phi\bigl(\mathsf{x}(\tau')\bigr)
\phi\bigl(\mathsf{x}(\tau'')\bigr)
\vert A\rangle
\ . 
\end{equation}

As the prefactor of $F(\omega)$ in (\ref{total-probability}) is a detector-specific constant that does depend on the trajectory of the detector or the state of the field, we shall work only with the response function $F(\omega)$ (and often abuse the language by calling it the transition probability). 

We take the initial state of the field to be the standard Minkowski vacuum, $\vert A\rangle=\vert 0\rangle$. The Wightman function $\langle 0\vert \phi(\mathsf{x})\phi(\mathsf{x}')\vert 0\rangle$ is then a well-defined distribution on $M^4 \times M^4$, and its pull-back to the detector world line, the correlation function 
\begin{equation}
\label{defW} 
W(\tau',\tau'') = 
\langle 0\vert 
\phi \bigl(\mathsf{x}(\tau') \bigr)
\phi \bigl(\mathsf{x}(\tau'') \bigr)
\vert 0\rangle
\ , 
\end{equation}
is a well-defined distribution on $\BbbR \times \BbbR$ \cite{Junker}. Thus, provided that the switching function $\chi(\tau)$ is smooth and of compact support, (\ref{defresponse}) gives an unambiguous answer to the question ``What is the probability of the detector being observed in the state $\vert1\rangle_d$ after the interaction has ceased?''

Following \cite{schlicht}, we introduce at this stage a convenient change of variables for the double integral over the $(\tau',\tau'')$ plane, making $u=\tau', s=\tau'-\tau''$ in the lower half-plane $\tau''< \tau'$ and $u=\tau'', s=\tau''-\tau'$ in the upper half-plane $\tau'< \tau''$. The response function becomes
\begin{equation}\label{newvariables}
 F(\omega)=2\,\mathrm{Re}\int_{-\infty}^{+\infty}\mathrm{d}u \,\chi(u)\int_{0}^{+\infty}\mathrm{d}s \,\chi(u-s)\, \mathrm{e}^{-i\omega s}\, W(u,u-s)\,.
\end{equation}

Though formulas (\ref{defresponse}) and (\ref{newvariables}) are suitable for calculating the detector's response, the results will not have a transparent separation between the properties of the response due to the trajectory and those due to the arbitrary switching functions; neither will they exhibit how the response depends on the proper time along the trajectory. Many authors \cite{schlicht,finitetime1,finitetime2,finitetime3}, therefore, prefer to ask instead the question: ``If the detector is turned on at time $\tau_0$ and read at time $\tau$, while the interaction is still on, what is the probability that the transition would have taken place?'' This amounts effectively to writing $\chi(\tau')=\Theta(\tau'-\tau_0)\Theta(\tau-\tau')$ (and similarly for $\chi(\tau'')$, $\chi(u)$ and $\chi(u-s)$) in the preceeding formulas. In this way we can talk of the transition probability as a function of the time $\tau$
\begin{equation}
F_{\tau}(\omega)=2\,\mathrm{Re}\int_{\tau_0}^{\tau}\mathrm{d}u\int_0^{u-\tau_0}\mathrm{d}s\,\mathrm{e}^{-i\omega s}\, W(u,u-s)\,,
\end{equation}
and define the instantaneous transition rate as its derivative with respect to $\tau$:
\begin{equation}\label{defexcitationrate}
\dot{F}_{\tau}(\omega)=2\,\mathrm{Re}\int_0^{\Delta\tau}\mathrm{d}s\,\mathrm{e}^{-i\omega s}\, W(\tau,\tau-s)\,,
\end{equation}
where $\Delta\tau=\tau-\tau_0$. This represents the number of excitations per unit time in an ensemble of identical detectors, up to a proportionality constant. When the trajectory is uniformly accelerated, and $\tau_0$ is taken to $-\infty$ to eliminate transient effects, this quantity should equal a Planckian thermal spectrum constant in $\tau$ according to the Unruh effect.

This procedure, however, creates a potential problem as to the definition of the Wightman function. It is proved in \cite{kay-wald} that the Wightman function is a Lorentz-invariant distribution, well-defined by the formula
\begin{equation}
 \label{covWightman}
\langle 0\vert \phi(\mathsf{x})\phi(\mathsf{x}')\vert 0\rangle
=
\lim_{\epsilon \to 0_+} 
\frac{-1}{4\pi^2}\frac{1}{{(t-t')}^2
- 
{\vert\mathbf{x}-\mathbf{x'}\vert}^2
- i \epsilon 
\bigl[ 
T(\mathsf{x}) - T(\mathsf{x}')
\bigr] 
- \epsilon^2}
\ , 
\end{equation}
where $T$ is \emph{any\/} global time function that increases to the future. The usual representation is obtained with the choice $T(\mathsf{x}) = t$ in a specific Lorentz frame, giving
\begin{equation}
\label{tradWightman}
\langle 0\vert \phi(\mathsf{x})\phi(\mathsf{x}')\vert 0\rangle
=
\lim_{\epsilon \to 0_+} 
\frac{-1}{4\pi^2}\frac{1}{{(t-t'-i\epsilon)}^2
- 
{\vert\mathbf{x}-\mathbf{x'}\vert}^2}
\ ,
\end{equation}
but the results obtained after integrating against functions of $x$ and $x'$ and taking the $\epsilon\rightarrow 0$ limit are independent of this choice. This is only garanteed to be true, however, as long as the functions integrated against are smooth, which is not the case when calculating an instantaneous transition rate.

The upshot of all this is that we have no reason to expect (\ref{defexcitationrate}) to give an unambiguous answer independent of the global time function $T$ used in the regularisation of the correlation function. In fact, Schlicht showed \cite{schlicht,Schlicht:thesis} that for the usual choice (\ref{tradWightman}) unphysical results were obtained for the uniformly accelerated trajectory, whereas the correct Plackian spectrum is obtained if instead of it we used the regularisation
 \begin{equation}
\label{corrSchlicht}
W_\epsilon (\tau,\tau')
=\lim_{\epsilon \to 0_+} 
\frac{1}{4\pi^2}\frac{1}{\bigl( \mathsf{x}-\mathsf{x}'
-i\epsilon 
(\dot{\mathsf{x}}+\dot{\mathsf{x}}') 
\bigr)^2}
\ . 
\end{equation}
Here $\mathsf{\dot{x}}$ and $\mathsf{\dot{x'}}$ are the four-velocity of the detector evaluated at $\tau$ and $\tau'$ respectively. This expression was obtained from a model of a spatially extended detector, in which the pointlike coupling to $\phi(\mathsf{x(\tau)})$ in (\ref{Hint}) is replaced by a coupling to a smeared field $\phi_f(\tau)$. The smearing is done integrating the field over the instantaeous simultaneity plane of the detector (parametrized with Fermi-Walker coordinates) using a profile function defining a rigid ``shape'' for the detector; this profile function is chosen by Schlicht to be a Lorentzian function with a size parameter $\epsilon$, which when taken to zero recovers the pointlike limit; this is the regularisation parameter appearing in (\ref{corrSchlicht}). 

In \cite{us} it is shown explicitly how (\ref{tradWightman}) and (\ref{corrSchlicht}) give different answers for a generic trajectory. Starting with the expression
\begin{equation}
\dot{F}_{\tau}(\omega)=\lim_{\epsilon \to 0_+} 2\,\mathrm{Re}\int_0^{\Delta\tau}\mathrm{d}s\,\mathrm{e}^{-i\omega s} W_\epsilon(\tau,\tau-s)\,,
\end{equation}
the limit $\epsilon\rightarrow 0$ is taken by dividing the integral in two intervals, $(0, \sqrt{\epsilon})$ and $(\sqrt{\epsilon},\Delta\tau)$. In the second interval the limit can be taken under the integral, while in the first one the integrand is expanded for small $s$ to a suitable power and then calculated exactly neglecting terms of order $\epsilon$. The complete result found for the excitation rate when Schlicht's modified correlation function (\ref{corrSchlicht}) is used is
\begin{equation}
\label{resultado1}
\dot{F}_{\tau}(\omega)
=
-\frac{\omega}{4\pi}+\frac{1}{2\pi^2}
\int_0^{\Delta\tau}\textrm{d}s
\left( 
\frac{\cos (\omega s)}{{(\Delta \mathsf{x})}^2} 
+ 
\frac{1}{s^2} 
\right) 
\ \ +\frac{1}{2\pi^2 \Delta \tau}
\ , 
\end{equation}
while when using instead the conventional correlation function given by (\ref{tradWightman}) we obtain
\begin{eqnarray}
\label{resultado2}
\lefteqn{
\dot{F}_{\tau}(\omega)
=
-\frac{\omega}{4\pi}+\frac{1}{2\pi^2}
\int_0^{\Delta\tau}\textrm{d}s
\left( 
\frac{\cos (\omega s)}{{(\Delta \mathsf{x})}^2} 
+ 
\frac{1}{s^2} 
\right) 
\ \ +\frac{1}{2\pi^2 \Delta \tau} } \nonumber\\
& & -\frac{1}{4\pi^2}\frac{\ddot{t}}{{(\dot{t}^2-1)}^{3/2}}
\left[ \dot{t}\sqrt{\dot{t}^2-1}
+\ln\!\left(\dot{t}-\sqrt{\dot{t}^2-1}\,\right)\right]
\ . 
\end{eqnarray}
(In both expressions, $(\Delta \mathsf{x})^2= (\mathsf{x}(\tau)-\mathsf{x}(\tau-s))^2$.)

We see that (\ref{resultado2}) is equal to (\ref{resultado1}) plus an extra term, which is independent of $\omega$ and non-Lorentz invariant (except for inertial trajectories, in which $\dot{t}=1$ and the term takes its limiting value $0$). It is this term that is responsible for the unphysical results found by Schlicht; if the excitation rate for a uniformly accelerated trajectory is calculated from (\ref{resultado1}) (setting $\Delta\tau=\infty$), the spectrum comes out Planckian. 

It seems reasonable to consider (\ref{resultado1}) to be the ``correct'' expression for the transition rate of a detector in arbitrary motion. It is manifestly causal and Lorentz invariant, it contains no regularisation parameters (resolving instead the pole at $s=0$ with the counterterm $s^{-2}$), it can be derived from a physical model of an extended detector in the zero-size limit, and it gives physically sensible results for a variety of trajectories \cite{us}. But we are left with the question to explain, why does the result for the detector that uses the conventional regularisation differ by the strange extra term in (\ref{resultado2})? We know it must have something to do with the sharp switching that violates the assumptions under which (\ref{covWightman}) works, but can we say any more about that?

In the next section we shall provide an answer to these questions. Starting from (\ref{newvariables}), we shall repeat the calculations to take the explicit $\epsilon\rightarrow 0$ limit but now with smooth switching functions in the picture, and we shall show that the Lorentz-noninvariant term does not appear. After that we will also show that when the switching functions approach step functions in a suitably controlled sense, the resulting transition rate approaches (\ref{resultado1})  

\section{Transition probability with smooth switching functions}

From (\ref{newvariables}), the response function for the pointlike detector with switching functions and using the conventional $i \epsilon$-regulated correlation function is
\begin{align}
 F(\omega) &= \frac{1}{2\pi^2}\lim_{\epsilon\rightarrow 0}
\int_{-\infty}^{+\infty}\mathrm{d}u\,\chi(u)
\int_0^{+\infty}\mathrm{d}s\,\chi(u-s)\,\times \nonumber\\
& \quad \Bigg[ \frac{\cos(\omega s)\left( (\Delta \mathsf{x}\right) ^2+\epsilon^2))}{\left( (\Delta \mathsf{x})^2+\epsilon^2)\right) ^2+4\epsilon^2\Delta t^2} 
 -\frac{\sin(\omega s)2\epsilon \Delta t}{((\Delta \mathsf{x})^2+\epsilon^2))^2+4\epsilon^2\Delta t^2}\Bigg]\,, 
\end{align}
with $\Delta \mathsf{x}= \mathsf{x}(u)-\mathsf{x}(u-s)$ and $\Delta t=t(u)-t(u-s)$. The switching function $\chi(\tau)$ is smooth and of compact support, and the trajectory $\mathsf{x}(\tau)$ is assumed to be $C^9$ in the interval where $\chi(\tau)$ does not vanish. We separate the $s$ integral in two intervals $(0,\eta)$ and $(\eta, +\infty)$, with $\eta=\sqrt{\epsilon}$, and for each integral consider separately the terms even and odd in $\omega$; thus the $s$ integral is broken in $I_<^{\mathrm{odd}}$, $I_<^{\mathrm{even}}$, $I_>^{\mathrm{odd}}$ and $I_>^{\mathrm{even}}$. Taking first the term $I_>^{\mathrm{even}}$, for $\epsilon=0$ the integrand reduces to $\chi(u-s)\cos(\omega s)/(\Delta \mathsf{x})^2 $, and the error involved in this replacement can be written:

\begin{equation}\label{even}
\int_\eta^{+\infty}\mathrm{d}s\,\chi(u-s) \cos(\omega s)
\frac{\epsilon^2}{\left[ (\Delta \mathsf{x})^2\right]^2 }
\frac{1+4\frac{(\Delta t)^2}{(\Delta \mathsf{x})^2}+\frac{\epsilon^2}{(\Delta \mathsf{x})^2}}
{\left\lbrace \left( 1+\frac{\epsilon^2}{(\Delta \mathsf{x})^2}\right)^2+4\epsilon^2\frac{(\Delta t)^2}{\left[ (\Delta \mathsf{x})^2\right]^2 } \right\rbrace} \,.
\end{equation}

From the small $s$ expansions of $(\Delta \mathsf{x})^2=-s^2+O(s^4)$ and $\Delta t=O(s)$ it follows that the quantities $(\Delta t)^2/(\Delta \mathsf{x})^2$ and $\epsilon/(\Delta \mathsf{x})^2$ are bounded by constants independent of $\epsilon$ over the interval of integration. Since $\bigl\vert {(\Delta\mathsf{x})}^2 \bigr\vert \ge s^2$, the absolute value of the integrand in (\ref{even}) is thus bounded by a constant times $\epsilon^2/s^4$ and the integral is of order $O(\eta)$. A similar estimate shows $I_>^{\mathrm{odd}}= O(\eta)$, and thus we have
\begin{equation}\label{2ndint}
 I_>^{\mathrm{even}}+I_>^{\mathrm{odd}} =
\int_{\eta}^{+\infty}\mathrm{d}s\,\chi(u-s)\frac{\cos(\omega s)}{(\Delta \mathsf{x})^2}\,\,+O(\eta).
\end{equation}

In the $(0,\eta)$ interval we expand the interval for small $s$ to obtain an estimate, following the similar calculation in \cite{us} where the details can be found. To control the estimates the quantity $(\Delta \mathsf{x})^2$ must be expanded to order $s^8$, which is why we assumed the trajectory to be $C^9$. Making the change of variables $s=\eta^2 x$ we obtain: 

\begin{subequations}\label{expansion}
\begin{align}
\label{expeven}
 I_<^{\mathrm{even}} &=
\frac{1}{\eta^2}\int_0^{1/\eta}\mathrm{d}x\,\frac{(\chi-\dot{\chi}\eta^2 x)(1-x^2)}{1+x^4+2x^2(2\dot{t}^2-1)}\left[ 1+\frac{4\dot{t}\ddot{t}\eta^2 x^3}{1+x^4+2x^2(2\dot{t}^2-1)}\right]\, +O(\eta) \,, \\
\label{expodd}
I_<^{\mathrm{odd}} &=
-\int_0^{1/\eta}\mathrm{d}x\,\frac{2\chi\omega x}{1+x^4+2x^2(2\dot{t}^2-1)}\,+O(\eta)\,.
\end{align}
\end{subequations}
Here $\chi$ and $t$ are evaluated at $u$, and a dot indicates differentiation with respect to $u$. The $O(\eta)$ estimate for the term outside the integral includes functions of $u$, which (because of the outer integral in $u$ with a $\chi(u)$ of compact support) are bounded by constants, ensuring that the estimate holds uniformly in $u$.

The integrals in (\ref{expansion}) are elementary, and joining them with (\ref{2ndint}) we obtain
\begin{align}
 F(\omega) &=\lim_{\eta\rightarrow 0} \frac{1}{2\pi^2}\int_{-\infty}^{+\infty}\mathrm{d}u\,\chi(u)
\Bigg[ \frac{\chi}{\eta}-\dot{\chi}\ln(\eta)-\frac{\chi\omega\pi}{2} 
+\frac{\dot{\chi}\dot{t}}{(\dot{t}^2-1)^{1/2}}\ln(\dot{t}-(\dot{t}^2-1)^{1/2})
\nonumber\\
& -\frac{\chi\ddot{t}}{2(\dot{t}^2-1)^{3/2}} \big[\dot{t}(\dot{t}^2-1)^{1/2}+\ln(\dot{t}-(\dot{t}^2-1)^{1/2})\big]
+\int_\eta^{+\infty}\mathrm{d}s\,\chi(u-s)\frac{\cos(\omega s)}{(\Delta \mathsf{x})^2}
\Bigg] \,,
\end{align}
with the terms involving $\dot{t}$ reducing to their limiting values in the $\dot{t}=1$ inertial case.

The term proportional to $\ln(\eta)$ vanishes by integration by parts using the fact that $\chi(u)$ is smooth and of compact support. The terms involving non-Lorentz invariant functions of the trajectory also vanish in the same way, because
\begin{displaymath}
 \frac{\mathrm{d}}{\mathrm{d}u}\left[ \frac{\dot{t}}{(\dot{t}^2-1)^{1/2}}\ln(\dot{t}-(\dot{t}^2-1)^{1/2})\right]=- \frac{\ddot{t}}{(\dot{t}^2-1)^{3/2}} \big[\dot{t}(\dot{t}^2-1)^{1/2}+\ln(\dot{t}-(\dot{t}^2-1)^{1/2})\big] \,,
\end{displaymath}
and therefore they cancel out when integrating by parts. Finally, the term $\chi(u)/\eta$ can be rewritten as part of the $s$-integral as follows:

\begin{equation}\label{inside-integral}
 F(\omega)=\lim_{\eta\rightarrow 0} \frac{1}{2\pi^2}\int_{-\infty}^{+\infty}\mathrm{d}u\,\chi(u)
\left[ -\frac{\chi(u)\omega\pi}{2} +\int_\eta^{+\infty}\mathrm{d}s\,\left( \chi(u-s)\frac{\cos(\omega s)}{(\Delta \mathsf{x})^2}+\frac{\chi(u)}{s^2}\right) \right] \,.
\end{equation}

We see that the non-Lorentz invariant term that appeared in (\ref{resultado2}) has been cancelled, thanks to the smooth switching function, and the results begin to resemble the transition rate (\ref{resultado1}) found from the extended detector model. To take the $\eta\rightarrow 0$ limit we add and subtract within the $s$-integral in (\ref{inside-integral}) a term $\chi(u-s)/s^2$ and group the terms in the following way: 

\begin{align}
F(\omega) &=\lim_{\eta\rightarrow 0}\Bigg[
-\frac{\omega}{4\pi}\int_{-\infty}^{+\infty}\mathrm{d}u\,[\chi(u)]^2\,+
\frac{1}{2\pi^2} \int_{-\infty}^{+\infty}\mathrm{d}u\,\chi(u)\int_\eta^{+\infty}\mathrm{d}s\,\chi(u-s)\left( \frac{\cos(\omega s)}{(\Delta \mathsf{x})^2}+\frac{1}{s^2}\right)  \nonumber\\
&+\frac{1}{2\pi^2}\int_\eta^{+\infty}\frac{\mathrm{d}s}{s^2}\int_{-\infty}^{+\infty}\mathrm{d}u\,\chi(u)\left[ \chi(u)-\chi(u-s)\right] \Bigg]\,.
\end{align}
The interchange of integrals in the last term is justified by the absolute convergence of the double integral. The $\eta\rightarrow 0$ limit of the last term is well-behaved because the $u$-integral regarded as a function of $s$ has a Taylor expansion starting with an $O(s^2)$, and taking the $\eta\rightarrow 0$ limit under the $u$-integral in the middle term can be justified by a dominated convergence argument. Thus the final result for the response function after the regularisation parameter is taken to $0$ reads 
\begin{align}\label{probability}
 F(\omega) &= -\frac{\omega}{4\pi}\int_{-\infty}^{+\infty}\mathrm{d}u\,[\chi(u)]^2\,
+\frac{1}{2\pi^2}\int_{-\infty}^{+\infty}\mathrm{d}u\,\chi(u)\int_0^{+\infty}\mathrm{d}s\,\chi(u-s)\left( \frac{\cos(\omega s)}{(\Delta \mathsf{x})^2}+\frac{1}{s^2}\right) \nonumber \\
&+ \frac{1}{2\pi^2}\int_0^{+\infty}\frac{\mathrm{d}s}{s^2}\int_{-\infty}^{+\infty}\mathrm{d}u\,\chi(u)\left[ \chi(u)-\chi(u-s)\right].
\end{align}

To summarize, (\ref{probability}) gives the probability for an Unruh-DeWitt detector to have made a transition of energy $\omega$, after being smoothly switched on and later off while following an arbitrary $C^9$ trajectory $\mathsf{x}(\tau)$ interacting with a massless scalar field originally in its vacuum state. We note that the broad features of (\ref{probability}) agree with those of the transition rate (\ref{resultado1}) found in \cite{us} from the the model with sharp swicthing and regularisation by spatial smearing. Both have a trajectory-independent term odd in $\omega$, a trajectory-dependent term even in $\omega$, and a term independent of both $\omega$ and the trajectory and which appears to have problems in the sharp switch-on limit. In (\ref{probability}) the problem is that the limit is ambiguous; in (\ref{resultado1}) it is that the transition probability obtained by integration is divergent. In the next section we show how to define a transition rate for sharp switching as a limiting case of a derivative of (\ref{probability}), in a way that agrees exactly with (\ref{resultado1}).

\section{The sharp switching limit}

We assume now the switching function $\chi(u)$ to take the form
\begin{equation}\label{chihs}
 \chi(u)=h_1\left( \frac{u-\tau_0-\delta}{\delta}\right)\times h_2\left( \frac{-u+\tau+\delta}{\delta}\right)
\end{equation}
with $\tau>\tau_0$, $\delta>0$, and $h_1(x)$ and $h_2(x)$ smooth functions satisfying $h_{1,2}=0$ for $x<0$ and $h_{1,2}=1$ for $x>1$. This means that the detector is turned on smoothly according to the function $h_1(x)$ during the interval $(\tau_0-\delta,\tau_0)$, remains turned on for an interval $\Delta\tau=\tau-\tau_0$, and is turned off smoothly according to the function $h_2(1-x)$ during the interval $(\tau,\tau+\delta)$. The sharp switching limit will thus be given by $\delta/\Delta\tau \rightarrow 0$.

Let us examine the behaviour of our result (\ref{probability}) in this limit. The first term reduces to

\begin{equation}
 -\frac{\omega}{4\pi}\int_{-\infty}^{+\infty}\mathrm{d}u\,[\chi(u)]^2= -\frac{\omega}{4\pi}\Delta\tau+O\left(\frac{\delta}{\Delta\tau}\right)
\end{equation}
The second term reduces in a similar unambiguous way (we omit the $1/2\pi^2$ prefactor):

\begin{multline}
\int_{-\infty}^{+\infty}\mathrm{d}u\,\chi(u)\int_0^{+\infty}\mathrm{d}s\,\chi(u-s)\left( \frac{\cos(\omega s)}{(\Delta \mathsf{x})^2}+\frac{1}{s^2}\right)=\\ \int_{\tau_0}^{\tau}\mathrm{d}u\,\int_0^{u-\tau_0}\mathrm{d}s\,\left( \frac{\cos(\omega s)}{(\Delta \mathsf{x})^2}+\frac{1}{s^2}\right)+O\left(\frac{\delta}{\Delta\tau}\right) \,.
\end{multline}

But we encounter a problem in the third term, which is divergent in the limit under consideration. Naive substitution of $\chi(u)$ by $\Theta(u-\tau_0)\Theta(\tau-u))$ yields
\begin{equation}\label{thirdterm}
 \int_0^{+\infty}\frac{\mathrm{d}s}{s^2}\int_{-\infty}^{+\infty}\mathrm{d}u\,\chi(u)\left[ \chi(u)-\chi(u-s)\right]=\int_0^{\Delta\tau}\frac{\mathrm{d}s}{s}\,,
\end{equation}
showing the divergence to be logarithmic. The presence of logarithmic divergences in the response function due to a sharp switching of the detector was analysed in \cite{finitetime2,finitetime3}. It is also implied by the term proportional to $1/\Delta\tau$ in expressions (\ref{resultado1}) or (\ref{resultado2}) for the transition rate. We will show presently that though the transition probability diverges in the sharp switching limit, a finite transition rate can be defined in this limit, and it would make a good aproximation to the actual number of transitions per unit time for observation times $\Delta\tau$ much longer than the switching time $\delta$ (but still short enough that the first-order approximation in perturbation theory applies).

We substitute the form given by (\ref{chihs}) in the left hand side of (\ref{thirdterm}) and perform the redefinitions $x=(u-\tau_0+\delta)/\delta$, $r=s/\delta$ and $b=(\Delta\tau+\delta)/\delta$, which show explicitly that the result (for fixed shape functions $h_{1,2}$) depends only on the parameter $b$:

\begin{align}
 &\int_0^{+\infty}\frac{\mathrm{d}s}{s^2}\int_{-\infty}^{+\infty}\mathrm{d}u\,\chi(u)\left[ \chi(u)-\chi(u-s)\right]= 
\int_0^{+\infty}\frac{\mathrm{d}r}{r^2}
\int_{\infty}^{+\infty}\mathrm{d}v\, h_1(x)h_2(b+1-v) \nonumber\\
 & \times \Bigg[h_1(v)h_2(b+1-v)- h_1(v-r)h_2(b+1-v+r)\Bigg]\,. 
\end{align}

To evaluate this expression, we separate the $r$-integral in five subintegrals for the intervals $(0,1)$, $(1,b-1)$, $(b-1,b)$, $(b,b+1)$ and $(b+1,+\infty)$. which we label $I_{1,2,3,4,5}$ respectively. In each of these only the range $(0,b+1)$ of the $v$-integral can make a contribution. For $I_1$ we have
\begin{align}
I_1 
&= \int_0^{+1}\frac{\mathrm{d}r}{r^2}\Bigg[\int_0^r\mathrm{d}v\,[h_1(v)]^2
+\int_r^1\mathrm{d}v\,h_1(v)[h_1(v)-h_1(v-r)]
+\int_b^{b+r}\mathrm{d}v\,\Big[ h_2(b+1-v) \nonumber\\
&
\times\left( h_2(b+1-v)-1\right) \Big] +
\int_{b+r}^{b+1}\mathrm{d}v\,h_2(b+1-v)[h_2(b+1-v)-h_2(b+1-v+r)]
\Bigg]
\end{align}
and now by shifting in the last two integrals $v\rightarrow v-b$ we see that $I_1$ is a constant independent of $b$. For $I_2$ we have
\begin{align}
I_2 
&= \int_1^{b-1}\frac{\mathrm{d}r}{r^2}\Bigg[\int_0^1\mathrm{d}v\,[h_1(v)]^2
+r-1+\int_r^{r+1}\mathrm{d}v\,[1-h_1(v-r)] \nonumber\\
&
+\int_b^{b+1}\mathrm{d}v\, h_2(b+1-v)[ h_2(b+1-v)-1] 
\Bigg]
\end{align}
which after redefining $v\rightarrow v-r$ and $v\rightarrow -v+b+1$ in the next-to-last and last integral respectively reduces to
\begin{equation}
 I_2=\ln(b-1)+\left(1- \frac{1}{b-1}\right) \int_0^1\mathrm{d}v\,\left[[h_1(v)]^2+[h_2(v)]^2-h_1(v)-h_2(v)\right] \,.
\end{equation}
In a similar way analysis of the remining terms shows that
\begin{subequations}
\begin{align}
I_3 &= I_4=\frac{1}{b}+O\left( \frac{1}{b^2}\right) \\
I_5 &= 1+\frac{1}{b}\left[-2+ \int_0^1\mathrm{d}v\,\left[[h_1(v)]^2+[h_2(v)]^2\right] \right] +O\left( \frac{1}{b^2}\right) \,.
\end{align} 
\end{subequations}
Thus the total response function, when the switching function is given by (\ref{chihs}) in the $\delta/\Delta\tau \rightarrow 0$ limit, is accurately given by
\begin{align}\label{responselimit}
 F(\omega) &=
\frac{\omega}{4\pi}\Delta\tau
+\frac{1}{2\pi^2}\int_{\tau_0}^{\tau}\mathrm{d}u\,\int_0^{u-\tau_0}\mathrm{d}s\,\left( \frac{\cos(\omega s)}{(\Delta \mathsf{x})^2}+\frac{1}{s^2}\right)+ \frac{1}{2\pi^2}\ln\left( \frac{\Delta\tau}{\delta}\right) \nonumber\\
& + C +O\left( \frac{\delta}{\Delta\tau}\right) \,
\end{align}
where $C$ is a constant independent of $\omega$, $\Delta\tau$ and $\delta$. The logarithmic divergence has been rendered explicit in the third term. Defining the transition rate by the derivative of (\ref{responselimit}) with respect to $\tau$ gives
\begin{equation}
\label{resultado1-sec4}
\dot{F}_{\tau}(\omega)
=
-\frac{\omega}{4\pi}+\frac{1}{2\pi^2}
\int_0^{\Delta\tau}\textrm{d}s
\left( 
\frac{\cos (\omega s)}{{(\Delta \mathsf{x})}^2} 
+ 
\frac{1}{s^2} 
\right) 
\ \ +\frac{1}{2\pi^2 \Delta \tau} + O\left( \frac{\delta}{(\Delta\tau)^2}\right) 
\, . 
\end{equation}

We see that the transition rate is well defined in the sharp switching limit, even if the transition probability is not. Moreover the limiting value of (\ref{resultado1-sec4}) agrees exactly with the expression (\ref{resultado1}) derived from the spatial profile model. As remarked in section 2, from this expression in the $\Delta\tau\rightarrow\infty$ limit one can find the Planckian spectrum for a uniformly accelerated trajectory.

However, as the next section will show, this result must be handled with great care. It holds for fixed $\omega$, but does not allow us to compute reliably the asymptotic behaviour for large $\omega$. In general, we need to remember that the detector only has a finite response function if switched smoothly, and that (\ref{resultado1-sec4}) holds only as a valid approximation to the excitation rate for fixed $\omega$ when $\Delta\tau\gg\delta$.

\section{Asymptotic falloff properties}

Let us go back to (\ref{probability}) to examine how the excitation probability behaves for large $\omega$. By adding and subtracting within the $s$ integral a term $\cos(\omega s)/(-s^2)$, given that ${(\Delta \mathsf{x})}^2=-s^2$ for an inertial trajectory, we can write the response function as
\begin{equation}\label{ineracc}
 F(\omega)=F_{\mathrm{in}}(\omega) + \frac{1}{2\pi^2}\ \int_{-\infty}^{+\infty}\mathrm{d}u\,\chi(u)\int_0^{+\infty}\mathrm{d}s\,\chi(u-s) \cos(\omega s)\left( \frac{1}{(\Delta \mathsf{x})^2}+\frac{1}{s^2}\right) \,,
\end{equation}
where $F_{\mathrm{in}}(\omega)$ is the response for inertial motion with switching function $\chi$. For a detector in eternal motion $F_{\mathrm{in}}(\omega)$ would vanish for positive $\omega$, but the model we are considering has excitations due to the switching even in inertial motion. If we define the sharp switching limit and the transition rate as in the previous section, then we have the well-known result:

\begin{equation}
 \dot{F}_{\tau\,\mathrm{in}}(\omega)=-\frac{\omega}{2\pi}\Theta(-\omega) 
\end{equation}
in the $\Delta\tau\rightarrow\infty$ limit.

Turning our attention to the second, noninertial term in (\ref{ineracc}), we notice that it is even in $\omega$, implying that the effects of any acceleration are to induce upwards and downwards transitions in the detector with equal probability. Physically we would expect this term to die off for large $\vert\omega\vert$ if the trajectory is smooth enough, as high frequency modes do not ``see'' the acceleration. We prove this by using the following theorem \cite{wong}:

If the function $h$ is $C^\infty$ in $[a,\infty)$
and $h^{(n)}(s)=O(s^{-1-\epsilon})$ as $s\rightarrow\infty$ for some
$\epsilon>0$ and every $n\geq0$, then
\begin{equation}
\label{theorem}
\int_a^\infty\mathrm{d}s \, h(s)
\mathrm{e}^{ixs}
\sim 
\mathrm{e}^{iax}\sum_{n=0}^\infty h^{(n)}(a)
\left( \frac{i}{x}\right)^{n+1} 
\quad\quad 
\mathrm{as}\,\,x\rightarrow\infty
\ .
\end{equation}

To apply this theorem we first interchange the $u$ and $s$ integrals in (\ref{ineracc}) (which is justified by absolute convergence of the double integral) and then write $\cos(\omega s) = \mathrm{Re}(\mathrm{e}^{i\omega s})$, $a=0$, and
\begin{equation}\label{defh}
h(s)=\int_{-\infty}^{+\infty}\mathrm{d}u\,\chi(u)\chi(u-s)\left( \frac{1}{(\Delta \mathsf{x})^2}+\frac{1}{s^2}\right)\,. 
\end{equation}
Assuming the trajectory to be $C^\infty$, we obtain from (\ref{theorem}) that for large $\omega$ noninertial terms in the spectrum go as
\begin{equation}\label{assympt}
 F_\mathrm{nonin}(\omega)\sim\frac{1}{2\pi^2}\sum_{n=1}^\infty \frac{(-1)^n}{\omega^{2n}}h^{(2n-1)}(0)\,.
\end{equation}

However, we can easily prove that $h(s)$ is even and therefore that all the odd derivatives appearing in (\ref{assympt}) vanish:

\begin{align}
 h(-s) & =
\int_{-\infty}^{+\infty}\mathrm{d}u\,\chi(u)\chi(u+s)\left( \frac{1}{[\mathsf{x}(u)-\mathsf{x}(u+s)]^2}+\frac{1}{s^2}\right)
\notag
\\
& = \int_{-\infty}^{+\infty}\mathrm{d}u\,\chi(\tilde{u}-s)\chi(\tilde{u})\left( \frac{1}{[\mathsf{x}(\tilde{u}-s)-\mathsf{x}(\tilde{u})]^2}+\frac{1}{s^2}\right)
\notag
\\
& = h(s)
\end{align}
where we have in the second line defined $\tilde{u}=u+s$ and in the third one use the symmetry of $(\Delta\mathsf{x})^2$ between two points. The conclusion is that the effects of acceleration in the response die off faster than any power of $\omega$ for large $\vert\omega\vert$.

Does this conclusion still hold when the switching is done sharply? There is no problem in evaluating the noninertial contribution to the response in the limit considered in the previous section, as the logarithmic divergence appears in the trajectory-independent term. When we take $\chi(u)\rightarrow \Theta(u-\tau_0)\Theta(\tau-u)$, the second term in (\ref{ineracc}) becomes
\begin{equation}\label{ineracc2}
 F_\mathrm{nonin}(\omega)=\frac{1}{2\pi^2} \int_0^{\Delta\tau}\mathrm{d}s\,\cos(\omega s)\int_{\tau_0+s}^{\tau}\mathrm{d}u\,\left( \frac{1}{(\Delta \mathsf{x})^2}+\frac{1}{s^2}\right) \,.
\end{equation}
Using a more general version of the theorem cited above, which follows from the Riemann-Lebesgue lemma (see \cite{wong} for details), the asymptotic behaviour of this is seen to be
\begin{equation}\label{assympt2}
 F_\mathrm{nonin}(\omega)\sim\frac{1}{2\pi^2}\mathrm{Re}\sum_{n=1}^\infty \left(\frac{i}{\omega}\right)^{n+1} \left[\tilde{h}^{(n)}(0)-\mathrm{e}^{i\omega\Delta\tau}\tilde{h}^{(n)}(\Delta\tau)\right]\,,
\end{equation}
with
\begin{equation} \label{defttilde}
 \tilde{h}(s)=\int_{\tau_0+s}^{\tau}\mathrm{d}u\,\left( \frac{1}{(\Delta \mathsf{x})^2}+\frac{1}{s^2}\right)\,.
\end{equation}
For a generic trajectory this implies a quadratic falloff; the leading order is
\begin{equation}\label{assympt3}
 F_\mathrm{nonin}(\omega)\sim\-\frac{1}{2\pi^2\omega^2}\left[\frac{(\ddot{\mathsf{x}}(\tau))^2}{24}-\frac{3(\ddot{\mathsf{x}}(\tau_0))^2}{24}-\cos(\omega\Delta\tau)\left( \frac{1}{[ \mathsf{x}(\tau)-\mathsf{x}(\tau_0)]^2}+\frac{1}{\Delta\tau^2}\right) \right]+O\left(\omega^{-3}\right)\,.
\end{equation}
Taking now the $\tau$-derivative to get the asymptotic behaviour for the transition rate, and taking also the limit $\Delta\tau\rightarrow\infty$ to avoid transient effects, we arrive to a result already derived in \cite{us} by direct asymptotic expansion of the transition rate (\ref{resultado1-sec4})
\begin{equation}
\label{largeomega}
\dot{F}_{\tau}(\omega)
=
-\frac{\omega}{2\pi}\Theta(-\omega)
+\frac{\ddot{\mathsf{x}} \cdot \dddot{\mathsf{x}}}{24 \pi^2 \omega^2}
+O\left( \omega^{-4}\right) 
\quad\quad 
\mathrm{as}\,\, \vert\omega\vert \rightarrow\infty
\ . 
\end{equation}

This quadratic falloff of the spectrum is an artifact of the sharp switching assumption, just like the divergence in the transition rate for $\Delta\tau\rightarrow 0$. For fixed $\omega$ and $\Delta\tau\gg\delta$ the response function and the transition rate are given indeed by (\ref{responselimit}) and (\ref{resultado1-sec4}); but this cannot be turned into an asymptotic series giving (\ref{assympt3}) for the large $\omega$-behaviour. Rather, $\delta$ and $\Delta\tau$ ought to be kept finite to ensure that the definition of the detector always meets the conditions of the Wightman distribution (use of smooth, compact-supported $\chi$); the response thus obtained always dies faster than any power of $\omega$ for large $\vert\omega\vert$, and it is this response which may be examined in the $\Delta\tau\gg\delta$ limit if we wish so.

\section{Conclusions}

The motivation for this paper was to find the resolution to a puzzle implied by the results of \cite{schlicht,Langlois,us}. It was shown in those papers that the excitation rate of a particle detector regularised with the usual $i\epsilon$ prescription for the correlation function contains unphysical, non-Lorentz invariant terms. A new paradigm for the regularisation was introduced via spatial smearing, which leads to the Lorentz-invariant transition rate quoted in (\ref{resultado1}). It was conjectured in \cite{us} that the failure of the the conventional prescription was related to the sharp switching of the detector assumed in both papers, which violates the conditions under which integration against the Wightman distribution is garanteed to give an unambiguous answer. We set out here to investigate in more detail this conjecture by analysing the model with smooth switching.

We showed in section 3 that when the switch-on and switch-off are smooth, the Lorentz-noninvariant terms disappear in the $\epsilon\rightarrow 0$ limit, and the response function can be written in the manifestly Lorentz invariant form
\begin{align}\label{probability-conc}
 F(\omega) &= -\frac{\omega}{4\pi}\int_{-\infty}^{+\infty}\mathrm{d}u\,[\chi(u)]^2\,
+\frac{1}{2\pi^2}\int_{-\infty}^{+\infty}\mathrm{d}u\,\chi(u)\int_0^{+\infty}\mathrm{d}s\,\chi(u-s)\left( \frac{\cos(\omega s)}{(\Delta \mathsf{x})^2}+\frac{1}{s^2}\right)  \nonumber\\
& + \frac{1}{2\pi^2}\int_0^{+\infty}\frac{\mathrm{d}s}{s^2}\int_{-\infty}^{+\infty}\mathrm{d}u\,\chi(u)\left[ \chi(u)-\chi(u-s)\right]\,.
\end{align}
In this way we have confirmed that the Lorentz-noninvariance found in \cite{schlicht,Langlois,us} is due to the sharp switching assumption used in these papers, and that the conventional regularisation can be used for detector response calculations as long as the switching is done smoothly.

In section 4 we analysed the limit of (\ref{probability-conc}) to the sharp switching case, using a model in which the switching takes place over an interval $\delta$ much shorter than the total detection time $\Delta\tau.$. The response function in this limit is given by the logarithmically divergent expression (\ref{responselimit}),
\begin{align}\label{responselimit-conc}
 F(\omega) &=
\frac{\omega}{4\pi}\Delta\tau
+\frac{1}{2\pi^2}\int_{\tau_0}^{\tau}\mathrm{d}u\,\int_0^{u-\tau_0}\mathrm{d}s\,\left( \frac{\cos(\omega s)}{(\Delta \mathsf{x})^2}+\frac{1}{s^2}\right)+ \frac{1}{2\pi^2}\ln\left( \frac{\Delta\tau}{\delta}\right) \nonumber\\
& + C +O\left( \frac{\delta}{\Delta\tau}\right) \,
\end{align}
while the transition rate is given by the finite expression (\ref{resultado1-sec4}),
\begin{equation}
\label{resultado1-conc}
\dot{F}_{\tau}(\omega)
=
-\frac{\omega}{4\pi}+\frac{1}{2\pi^2}
\int_0^{\Delta\tau}\textrm{d}s
\left( 
\frac{\cos (\omega s)}{{(\Delta \mathsf{x})}^2} 
+ 
\frac{1}{s^2} 
\right) 
\ \ +\frac{1}{2\pi^2 \Delta \tau} + O\left( \frac{\delta}{(\Delta\tau)^2}\right) 
\ .
\end{equation}

This agrees with the results (\ref{resultado1}) found from the regularisation by spatial smearing. Nevertheless, it was also found in Section 5 that the asymptotic large $\vert\omega\vert$ behaviour is not given accurately from expansion of this limiting expression; naive expansion from (\ref{resultado1-conc}) gives a generically quadratic falloff for the noninertial terms whereas expansion from (\ref{probability-conc}) shows that these terms are in fact suppressed faster than any power of $\omega$. 

Sriramkumar and Padmanabhan \cite{finitetime3} have adressed the logarithmic divergence as related to the sharp switching by a different procedure, using Gaussian or exponential window functions to switch the detector. While their main results are consistent with ours, the approach used in section 4 of the present paper is more general and allows for separate control of the parameters $\delta$ and $\Delta\tau$. The calculations of \cite{finitetime3} also assume that the trajectory is stationary, while we have put no constraints on it beyond $C^9$ continuity.

An interesting feature of expression (\ref{probability-conc}) is that it distinctly shows in which ways the response function is affected by the switching functions and by the detector's trajectory; the latter is involved only in the second of the three terms, It would be interesting to compare this results with the method used by Davies and Ottewill \cite{davies}, who use a Fourier transform of the switching functions to isolate a contribution to the response that does not depend on them but only on the trajectory and the state of the field. Can the two appraches be related somehow? Another relevant comparison might be with approaches in which the detector's trajectory is dynamically determined, allowing for backreaction. \cite{casadio, hu}

The only advantage of the spatial profile regularisation over the more conventional $i\epsilon$ one used here is then that it manages to obtain the correct transition rate for sharp switching directly, without using a smooth switching function as an intermediate step. However, it is not clear whether this is a real asset, as the use of the switching function makes indisputable the validity of using the Wightman function as a distribution. Also the transition probability diverges for a sharply switched-on detector, although as discussed in \cite{us} the excitation probability remains finite for a detector in eternal motion with suitable asymptotics in the distant past. Moreover, the approach to detector response used in this paper is much easier to generalise to curved backgrounds than the spatial profile one, because the latter involves defining the rigid profile as a function of Fermi-Walker coordinates, something which may not be possible in a general setting. For these reasons using the conventional regularisation and bearing in mind that smooth switching functions must be included in the model to assure physical results seems a more recommendable policy.

\section*{Acknowledgements}

I wish to thank Jorma Louko for continued support, discussions and for reading the manuscript, and Bill Unruh for helpful suggestions. I am grateful for the hospitality of Department of Physics and Astronomy at the University of  British Columbia, where part of this work was done with the support of a Universitas 21 Prize Scholarship. This work was supported by an EPSRC Dorothy Hodgkin Research Award to the University of Nottingham.


\begin{thebibliography}{99}

\bibitem{unruh} W.~G. Unruh, Notes on black-hole evaporation, Phys.\ Rev.\ D \textbf{14}, 870 (1976).
\bibitem{deWitt} B.~S. DeWitt,``Quantum gravity, the new synthesis'', in \textit{General Relativity; an Einstein centenary survey\/} ed S.~W. Hawking and W.~Israel (Cambridge University Press, 1979) 
\bibitem{hawking} J.~B. Hartle and S.~W. Hawking, Path-integral derivation of black-hole radiance, Phys.\ Rev.\ D \textbf{13}, 2188 (1976).
\bibitem{gibb-haw:dS} G.~W. Gibbons and S.~W. Hawking, Cosmological event horizons, thermodynamics, and particle creation, Phys.\ Rev.\ D \textbf{15}, 2738 (1977). 
\bibitem{schlicht} S.~Schlicht, Considerations on the Unruh effect: causality and regularization, Class.\ Quantum Grav.\ \textbf{21} 4647 (2004). \lanln{gr-qc/0306022}
\bibitem{Schlicht:thesis} S.~Schlicht, ``Betrachtungen zum Unruh-Effekt: Kausalit\"at und Regularisierung", PhD Thesis, University of Freiburg (2002). 
\bibitem{Langlois} P.~Langlois, Causal particle detectors and topology, Ann.\ Phys.\ (N.Y.) \textbf{321} 2027
(2006). \lanln{gr-qc/0510049}
\bibitem{Langlois-thesis} P.~Langlois, ``Imprints of spacetime topology in the Hawking-Unruh effect", PhD Thesis, University of Nottingham (2005). \lanln{gr-qc/0510127}
\bibitem{us} J.~Louko and A.~Satz, How often does the Unruh–DeWitt detector click? Regularization by a spatial profile, Class.\ Quantum Grav.\ \textbf{23} 6321 (2006). \lanln{gr-qc/0606067}
\bibitem{kay-wald} B.~S. Kay and R.~M. Wald,  Theorems On The Uniqueness And Thermal Properties Of Stationary, Nonsingular, Quasifree States On Space-Times With A Bifurcate Killing Horizon, Phys.\ Rept.\ \textbf{207} 49, (1991). 
\bibitem{Junker} W.~Junker and E.~Schrohe,  Adiabatic vacuum states on general space-time manifolds: Definition, construction, and physical properties, Ann.\ Henri Poincar\'e \textbf{3}, 1113 (2002). \lanln{math-ph/0109010}
\bibitem{byd} N.~D. Birrell and P.~C.~W. Davies, \textit{Quantum Fields in Curved Space} (Cambridge University Press 1982). 
\bibitem{wald-smallbook} R.~M. Wald, \textit{Quantum field theory in curved spacetime and black hole thermodynamics\/} (University of Chicago Press, Chicago, 1994). 
\bibitem{finitetime1} B.~F. Svaiter and N.~F. Svaiter, Inertial and noninertial particle detectors and vacuum fluctuations, Phys.\ Rev.\ D \textbf{46} 5267 (1992).
\bibitem{finitetime2} A.~Higuchi, G.~E.~A. Matsas and C.~B. Peres, Uniformly accelerated finite-time detectors, Phys.\ Rev.\ D \textbf{48} 3731 (1993). 
\bibitem{finitetime3} L.~Sriramkumar and T.~Padmanabhan, Finite-time response of inertial and uniformly accelerated Unruh - DeWitt detectors, Class.\ Quantum Grav.\ \textbf{13} 2061 (1996). \lanln{gr-qc/9408037}
\bibitem{wong} R. Wong, \textit{Asymptotic Approximations of Integrals\/} (Society for Industrial and Applied Mathematics, Philadelphia, 2001). 
\bibitem{davies} P.~C.~W. Davies and A.~C. Ottewill, Detection of negative energy: 4-dimensional examples, Phys.\ Rev.\ D \textbf{65}, 104014 (2002) \lanln{gr-qc/0203003}
\bibitem{casadio} R. Casadio and G. Venturi, The accelerated observer with back-reaction effects, Phys.\ Lett.\ A \textbf{252} 109 (1999)
\bibitem{hu} Shih-Yuin Lin, B. L. Hu, Where is the Unruh Effect? - New Insights from Exact Solutions of Uniformly Accelerated Detectors (2006) \lanln{gr-qc/0611062}
\end{thebibliography}
\end{document}